\documentclass[letterpaper,aps,prb,twocolumn,amsmath,showpacs,amssymb]{revtex4}

\usepackage{epsfig}
\usepackage{graphicx}
\usepackage{dcolumn}
\usepackage{bm}
\usepackage{bbm}
\usepackage{wasysym}
\usepackage{marvosym}
\usepackage{times,amsmath,amssymb}

\newcommand{\trG}{\mathbbm G}

\newlength{\textwidthm}

\begin{document}

\title{The optical conductivity of graphene in the
visible region of the spectrum}

\author{T.~Stauber,$^1$ N.~M.~R.~Peres,$^1$ and A. K. Geim$^2$}

\affiliation{$^1$Centro de F\'{\i}sica  e  Departamento de
F\'{\i}sica, Universidade do Minho, P-4710-057, Braga, Portugal}

\affiliation{$^2$Manchester Centre for Mesoscience and Nanotechnology,
University of Manchester, Manchester M12 9PL, United Kingdom}

\date{\today}

\begin{abstract}
We compute the optical conductivity of graphene beyond the usual Dirac
cone approximation, giving results that are valid in the visible
region of the conductivity spectrum. The effect of next nearest
neighbor hoping is also discussed. Using the full expression for the
optical conductivity, the transmission and reflection coefficients are
given. We find that even in the optical regime the corrections to the
Dirac cone approximation are surprisingly small (a few percent). Our
results help in  the interpretation of the experimental results reported
by Nair {\it et al.} [Science {\bf 320}, 1308 (2008)].
\end{abstract}


\maketitle

\section{Introduction} 

Graphene, an atomically thin material made only of carbon atoms arranged
in a hexagonal lattice, was isolated only recently.\cite{Nov04,pnas}
Several reviews on the physics of graphene are already
available in the literature.\cite{Nov07,peresworld,rmp,rmpBeenakker}

At low energies, $E<1$ eV, the electronic dispersion has the form
$\epsilon(\bm k)=\pm 3tka/2$, where $t$ is the nearest neighbor
hopping integral and $a$ is the carbon-carbon distance. The effective
theory at these energy scales is that of a massless Dirac Hamiltonian
in (2+1) dimensions. If the experimental probes excite the system
within this energy range, the Dirac Hamiltonian is all there is for
describing the physics of graphene. On the other hand, for excitations
out of this energy range it is necessary to include corrections to the
Dirac Hamiltonian which will modify the energy spectrum and thus the
density of states of the system. One immediate consequence is that the
energy dispersion is no longer a function of the absolute value of the
wave-number $k$. In this paper, we will calculate the optical conductivity of graphene including the leading corrections to the Dirac cone approximation.

One of the first calculations of the optical conductivity of graphene, using
the Dirac Hamiltonian were done by Gusynin and
Sharapov.\cite{Gusynin06} This first study was subsequently revisited
a number of times, \cite{Gusynin06PRL,Gusynin07PRL,Gusynin07} and
summarized in Ref.  [\onlinecite{GusyninIJMPB}]. However, these
authors did not include non-linear effects in the calculation. Also
the effect of disorder was done on a phenomenological level, by
broadening the delta functions into Lorentzians characterized by
constant width $\Gamma$. We note that in the Dirac-cone approximation, the
conductivity can also be obtained from the polarization. The
calculations for finite chemical potential and arbitrary $|\bm q|$ and
$\omega$ were done by Wunsch {\it et al.}\cite{Wunsch06} and Hwang and
Das Sarma.\cite{Hwang07}

The calculation of the optical conductivity of graphene, in the Dirac
Hamiltonian limit, including the effect of disorder in a self
consistent way was done by Peres {\it et al.}\cite{PeresPRB} and
recently also corrections due to electron-electron interaction were
discussed.\cite{Herbut08,Sachdev08} The calculation for the graphene
bilayer with disorder was done by Koshino and Ando, \cite{Koshino07}
and by Nilsson {\it et al.}.\cite {Nilsson07} The optical conductivity
of a clean bilayer was first computed by Abergel and Falko,
\cite{Falko07}, and recently generalized to the
biased\cite{StauberFR,castroPRL,Morpurgo07} bilayer case by Nicol and
Carbotte.\cite{Nicol08}

Within the Boltzmann approach, the optical conductivity of graphene
was considered in Refs. [\onlinecite{PeresBZ,StauberBZ}], where the
effect of phonons and the effect of mid-gap states were included.
This approach, however, does not include transitions between the
valence and the conduction band and is, therefore, restricted to
finite doping. The voltage and the temperature dependence of the
conductivity of graphene was considered by Vasko and Ryzhii,
\cite{Vasko07} using the Boltzmann approach. The same authors have
recently computed the photoconductivity of graphene, including the
effect of acoustic phonons.\cite{Vasko08}

The effect of temperature on the optical conductivity of clean
graphene was considered by L. A. Falkovsky and
A. A. Varlamov.\cite{Falkovsky07} The far-infrared properties of clean
graphene were studied in Refs. \onlinecite{Falkovsky08} and
\onlinecite{FalkovskyB}. Also this study was restricted to the Dirac
spectrum approximation.

It is interesting to note that the conductivity of clean graphene, at
half filling and in the limit of zero temperature, is given by the universal
value $\pi e^2/(2h)$.  \cite{Ludwig94,PeresIJMPB} On the other hand,
if the temperature is kept finite the conductivity goes to zero at
zero frequency, but the effect of optical phonons does not change the
value of the conductivity of clean graphene.\cite{Stauber08} This
behavior should be compared with the calculation of the DC
conductivity of disordered graphene, which for zero chemical potential
presents the value of $4e^2/(\pi h)$.\cite{PeresPRB,Ludwig94,Shon,Ando2002}

From the experimental point of view, the work of Kuzmenko {\it et al.}
\cite{Kuzmenko} studied the optical conductivity of graphite in the
energy range [0,1] eV, and showed that its behavior is close to that
predicted for clean graphene in that energy range. An explanation of
this odd fact was attempted within the Slonczewski-McClure-Weiss
model. The complex dielectric constant of graphite was studied by
Pedersen for all energy ranges.\cite{Pedersen} The infrared
spectroscopy of Landau levels in graphene was studied by Jiang {\it et
al.}\cite{Jiang} and Deacon {\it et al.}\cite{Deacon}, confirming the
magnetic field dependence of the energy levels and deducing a band
velocity for graphene of $1.1\times 10^6$ m/s. Recently, the infrared
conductivity of a single graphene sheet was
obtained.\cite{basov,Peres08}

Recent studies of graphene multilayers grown on SiC from THz to
visible optics showed a rather complex behavior\cite{George} with
values of optical conductivity close to those predicted for graphene
at infrared frequencies as well as to those measured in
graphite\cite{Kuzmenko}. This experiment\cite{George} especially
indicates the need for a graphene theory valid all the way to optical
frequencies. The absorption spectrum of multilayer graphene in high
magnetic fields was recently discussed in Ref. \onlinecite{deHeer},
including corrections to the Dirac cone approximation.

In this paper we address the question of how the conductivity of clean
graphene changes when ones departs from the linear spectrum
approach. This is an important question for experiments done in the
visible region of the spectrum.\cite{nair} The paper is organized as
follows: in Sec. \ref{hamilt} we introduce our model and derive the
current operator; in Sec. \ref{OC} we discuss the optical conductivity
of graphene by taking into account its full density of states; in
Sec. \ref{stpprime} we discuss the effect on the optical conductivity
of a next nearest neighbors hopping term; in Sec. \ref{scattering} we
analyze the scattering of light by a graphene plane located at the
interface of two different dielectrics and give the transmissivity and
reflectivity curves in the visible region of the spectrum; finally in
Sec. \ref{Concl} we give our conclusions.

\section{The Hamiltonian and the current operators} 
\label{hamilt}

\begin{figure}[!ht]
\begin{center}
\includegraphics[scale=0.4]{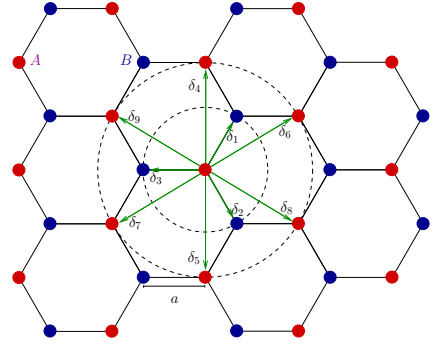}
\end{center}
\caption{
(color online) Representation of the vectors $\bm\delta_i$, with $i=1-9$.
The carbon-carbon distance, $a$, and the $A$ and $B$ atoms
are also depicted.
\label{hopping_tp}}
\end{figure}

The Hamiltonian, in tight binding form, for electrons in graphene
is written as
\begin{eqnarray}
H&=&-t\sum_{\bm R,\sigma}\sum_{\bm \delta=\bm \delta_1-\bm\delta_3}
[a^\dag_\sigma(\bm R)b_\sigma(\bm R+\bm \delta)+H.c.]\nonumber\\
&&-\frac {t'}2\sum_{\bm R,\sigma}\sum_{\bm \delta=\bm \delta_4-\bm\delta_9}
[a^\dag_\sigma(\bm R)a_\sigma(\bm R+\bm \delta)+H.c.]\nonumber\\
&&-\frac {t'}2\sum_{\bm R,\sigma}\sum_{\bm \delta=\bm \delta_4-\bm\delta_9}
[b^\dag_\sigma(\bm R)b_\sigma(\bm R+\bm \delta)+H.c.]\,,
\end{eqnarray} 
where the operator $a^\dag_\sigma(\bm R)$ creates an electron in the
carbon atoms of sub-lattice $A$, whereas $b^\dag_\sigma(\bm R)$
does the same in sub-lattice $B$, $t$ is the hopping parameter connecting 
first nearest neighbors, with a value
of the order of 3 eV, and $t'$ is the hopping parameter for second
nearest neighbors, with a value of the order of 0.1$t$.
The vectors $\bm\delta_i$ are represented in Fig. \ref{hopping_tp}
and have the form

\begin{equation}
\begin{array}{l}
\displaystyle{\boldsymbol{\delta}_1=\frac{a}{2}\left(1,\sqrt{3}\right)\;,\;
\boldsymbol{\delta}_2=\frac{a}{2}\left(1,-\sqrt{3}\right)\;,\;
\boldsymbol{\delta}_3= -a\left(1,0\right)}\;,\;\\ \\

\displaystyle{\boldsymbol{\delta}_4=a\left(0,\sqrt{3}\right)\;,\;\boldsymbol{\delta}_5=-\boldsymbol{\delta}_4\;,\;
\boldsymbol{\delta}_6=\frac{3a}{2}\left(1,\frac{1}{\sqrt{3}}\right)}\;,\; \\ \\

\displaystyle{\boldsymbol{\delta}_7=-\boldsymbol{\delta}_6\;,\;
\boldsymbol{\delta}_8=\frac{3a}{2}\left(1,-\frac{1}{\sqrt{3}}\right) \;,\;
\boldsymbol{\delta}_9=-\boldsymbol{\delta}_8}\;.
\end{array}
\end{equation}

In order to obtain the current operator we modify the 
hopping parameters as
\begin{equation}
t\rightarrow te^{i\frac{e}{\hbar}\bm A(t)\cdot\bm\delta}\,,
\end{equation}
and the same for $t'$. Expanding the exponential up to second order
in the vector potential $\bm A(t)$ and assuming that the electric
field is oriented along the $x$ direction, the current operator is obtained
from
\begin{equation}
j_x=-\frac{\partial H}{\partial A_x(t)}\,,
\end{equation}
leading to $j_x=j_x^P+A_x(t)j^D_x$. The operator $j_x^P$ reads
\begin{eqnarray}
j_x^P&=&\frac {tie}{\hbar}
\sum_{\bm R,\sigma}\sum_{\bm \delta=\bm \delta_1-\bm\delta_3}
[\delta_x a^\dag_\sigma(\bm R)b_\sigma(\bm R+\bm \delta)- H.c.]\nonumber\\
&+&\frac {t'ie}{2\hbar}\sum_{\bm R,\sigma}\sum_{\bm \delta=\bm \delta_4-\bm\delta_9}
[\delta_xa^\dag_\sigma(\bm R)a_\sigma(\bm R+\bm \delta)-H.c.]\nonumber\\
&+&\frac {t'ie}{2\hbar}\sum_{\bm R,\sigma}\sum_{\bm \delta=\bm \delta_4-\bm\delta_9}
[\delta_xb^\dag_\sigma(\bm R)b_\sigma(\bm R+\bm \delta)-H.c.]\,.
\end{eqnarray}
The operator $j^D_x$ can be found from the linear term in $A_x(t)$ expansion of
the Hamiltonian.
\section{The optical conductivity } 
\label{OC}
\subsection{The Kubo formula}

The Kubo formula for the conductivity is given by 
\begin{equation}
\sigma_{xx}(\omega) = \frac {< j^D_x>}{iA_s(\omega + i0^+)}+
\frac {\Lambda_{xx}(\omega + i0^+)}{i\hbar A_s(\omega + i0^+)}\,,
\end{equation}
with $A_s=N_cA_c$ the area of the sample, and $A_c=3\sqrt 3 a^2/2$ 
($a$ is the carbon-carbon distance)
the area of the unit cell,
from which it follows that
\begin{equation}
\Re\sigma_{xx}
(\omega) = D\delta(\omega) + \frac {\Im \Lambda_{xx}(\omega + i0^+)}
{\hbar\omega A_s}\,,
\end{equation}
and
\begin{equation}
\Im\sigma_{xx}
(\omega) = -\frac {< j^D_x>}{A_s\omega} - \frac {\Re \Lambda_{xx}(\omega + i0^+)}
{\hbar\omega A_s}\,,
\end{equation}

where $D$ is the charge stiffness which reads
\begin{equation}
D= -\pi \frac {<j^D_x>}{A_s} -\pi\frac {\Re \Lambda_{xx}(\omega + i0^+) }
{\hbar A_s}\,.
\label{DW}
\end{equation}
The function $\Lambda_{xx}(\omega + i0^+)$ is obtained from the 
Matsubara current-current correlation function, defined as
\begin{equation}
\Lambda_{xx}(i\omega_n) = \int_0^{\hbar\beta}d\,\tau e^{i\omega_n\tau}
<T_{\tau} j^P_{x}(\tau)j^P_x(0)>\,.
\end{equation}

In what follows we start by neglecting the contribution of $t'$ to the
current operator. Its effect is analyzed later and shown to be
negligible.  The function $\Im \Lambda_{xx}(\omega + i0^+)$ is given
by
\begin{eqnarray}
&&\Im \Lambda_{xx}(\omega + i0^+)=\frac {t^2e^2a^2}{8\hbar^2} \sum_{\bm k}
f[\phi(\bm k)] 
\nonumber\\
&\times&
[n_F(-t\vert\phi(\bm k)\vert-\mu)
-n_F(t\vert\phi(\bm k)\vert-\mu)]\nonumber\\
&\times&[\pi \delta (\omega -2t\vert\phi(\bm k)\vert/\hbar) -
\pi \delta (\omega +2t\vert\phi(\bm k)\vert/\hbar)
]
\,,
\label{im}
\end{eqnarray}
where $n_F(x)$ is the usual Fermi distribution, $\mu$ is the chemical potential,
and the function $\Re \Lambda_{xx}(\omega + i0^+)$ is given by 
\begin{eqnarray}
&&\Re \Lambda_{xx}(\omega + i0^+)=-\frac {t^2e^2a^2}{8\hbar^2} 
{\cal P}
\sum_{\bm k}
f[\phi(\bm k)] 
\nonumber\\
&\times&
[n_F(-t\vert\phi(\bm k)\vert-\mu)
-n_F(t\vert\phi(\bm k)\vert-\mu)]\nonumber\\
&\times&
\frac {4t\vert\phi(\bm k)\vert}{\omega^2- (2\vert\phi(\bm k)\vert)^2}
\,,
\label{Re}
\end{eqnarray}
with
\begin{equation}
\label{FormFactor}
f[\phi(\bm k)] = 18-4\vert\phi(\bm k)\vert^2 + 18 \frac {[\Re\phi(\bm k)]^2-[\Im \phi(\bm k)]^2}
{\vert\phi(\bm k)\vert^2}\,,
\end{equation}
and ${\cal P}$ denoting the principal part of the integral.
The graphene energy bands are given by 
$\epsilon(\bm k) = \pm t\vert\phi(\bm k)\vert$, with $\phi(\bm k)$
defined as
\begin{equation}
\phi(\bm k)=1 + e^{\bm k\cdot(\bm\delta_1-\bm\delta_3)}+
 e^{\bm k\cdot(\bm\delta_2-\bm\delta_3)}\,.
\end{equation}

\subsection{The real part of the conductivity}

The expression for (\ref{im}) can almost be written in terms of the
energy dispersion $\epsilon(\bm k)$, except for the term
\begin{equation}
\label{Neglect}
\frac {[\Re\phi(\bm k)]^2-[\Im \phi(\bm k)]^2}
{\vert\phi(\bm k)\vert^2}\,.
\end{equation}
In order to proceed analytically, and for the time being 
(see Section \ref{ugly}),
we approximate this term by
its value calculated in the Dirac cone approximation (see appendix A)
\begin{equation}  
\frac 1 {N_c}\sum_{\bm k}\frac {[\Re\phi(\bm k)]^2-[\Im \phi(\bm k)]^2}
{\vert\phi(\bm k)\vert^2}g(\vert\phi(\bm k)\vert)\simeq 0\,,
\label{Eq_dir}
\end{equation}
where $g(\vert\phi(\bm k)\vert)$ is some given function depending only
on the modulus
of $\phi(\bm k)$.
With this approximation, we have 
\begin{equation}
f[\phi(\bm k)] \simeq 18-4\vert\phi(\bm k)\vert^2\,.
\end{equation}
Introducing  the density of states per spin per unit cell, $\rho(E)$,
defined as
\begin{equation}
\rho(E)=\frac 1 {N_c}\sum_{\bm k}
\delta(E-t\vert\phi_{\bm k}\vert)\,,
\label{Eqrho}
\end{equation} 
the expression for the real part of the conductivity reads
\begin{eqnarray}
\Re\sigma_{xx}(\omega)=\sigma_0\frac {\pi
t^2a^2}{8A_c\hbar\omega}\rho(\hbar\omega/2) [18-(\hbar\omega)^2/t^2]
\nonumber\\ \times
\left[\tanh\frac{\hbar\omega+2\mu}{4k_BT}+\tanh\frac{\hbar\omega-2\mu}{4k_BT}
\right]\,.
\label{Eq_s}
\end{eqnarray}
Equation \ref{Eq_s} is essentially exact in the visible range of the spectrum; missing is only
the a contribution coming from Eq. \ref{Eq_dir}, which contribution wil later
be shown to be negligible.
In the above equation $\sigma_0$ is
\begin{equation}
\sigma_0=\frac \pi 2 \frac {e^2}h\,.
\end{equation}

\begin{figure}[t]
\begin{center}
\includegraphics*[angle=0,width=0.8\linewidth]{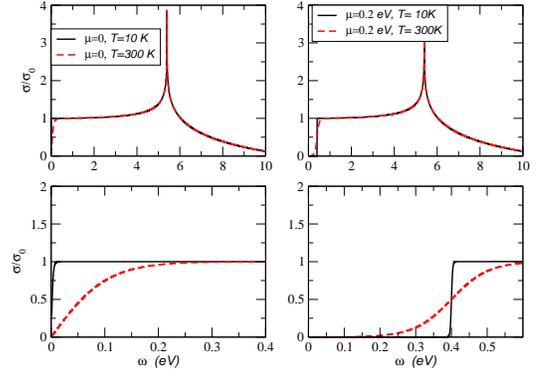}
\caption{(color online) The optical conductivity as function of 
frequency for two values of the chemical potential,
$\mu=0$ eV and $\mu=0.2$ eV, and two
temperatures $T=10$ K and $T=300$ K. 
The bottom panels are a zoom in, close to zero frequency, which allow to
depict the frequency region where differences in the chemical potential and in temperature
are most important. We have used $t=2.7$eV. 
\label{fig1}}
\end{center}
\end{figure}

The momentum integral in Eq. (\ref{Eqrho})
can be performed leading to
\begin{equation}
\rho(E)=\frac{2E}
{t^2\pi^{2}}\left\{ \begin{array}{ccc}
\frac{1}{\sqrt{F(E/t)}}\mathbf{K}
\left(\frac{4E/t}{F(E/t)}\right)\,, &  & 0<E<t\,,\\
\\\frac{1}{\sqrt{4E/t}}\mathbf{K}\left(\frac{F(E/t)}
{4E/t}\right)\,, &  & t<E<3t\,,\end{array}\right.
\label{eq:DOS1L}
\end{equation}  
where $F(x)$ is given by
\begin{equation}
F(x)=(1+x)^{2}-\frac{(x^{2}-1)^{2}}{4}\,,
\label{eq:Fe}
\end{equation}
and $\mathbf{K}(m)$ is defined as
\begin{equation}
\mathbf{K}(m)=\int^1_0 dx [(1-x^2)(1-mx^2)]^{-1/2}\,.
\label{eq:K}
\end{equation}
In Figure \ref{fig1} we give a plot of Eq. (\ref{Eq_s}) over
a large energy range, including the visible part of the spectrum
($E\in[1.0,3.1]$ eV).

It is useful to derive from Eq. (\ref{Eq_s}) an asymptotic expansion for $\Re\sigma_{xx}(\omega)$.
For that, we expand the density of states around $E=0$ and obtain
\begin{equation}
\rho(E)\simeq \frac {2E}{\sqrt 3 \pi t^2}+\frac {2E^3}{3\sqrt 3 \pi t^4}
+\frac {10E^5}{27\sqrt 3 \pi t^6}\,.
\label{expand_rho}
\end{equation}
Using Eq. (\ref{expand_rho}) in Eq. (\ref{Eq_s}) we obtain for
the optical conductivity the approximate result

\begin{eqnarray}
\Re\sigma_{xx}(\omega)&=&\sigma_0
\left(
\frac 1 2 +\frac 1 {72}\frac {(\hbar\omega)^2}{t^2}
\right)\nonumber\\
&\times&
\left(
\tanh\frac{\hbar\omega+2\mu}{4k_BT}+
\tanh\frac{\hbar\omega-2\mu}{4k_BT}
\right)\,.
\end{eqnarray}
In the case of $\mu=0$ this expression is the same
as in Kuzmenko {\it et al.}\cite{Kuzmenko}
 and in Falkovsky and Pershoguba
\cite{Falkovsky08}
if in both cases the $(\hbar\omega/t)^2$ term is neglected.
\subsection{Correction to $\Re\sigma_{xx}(\omega)$
introduced by Eq. (\ref{Eq_dir})}
\label{ugly}

\begin{figure}[t]
\begin{center}
  \includegraphics*[angle=0,width=0.8\linewidth]{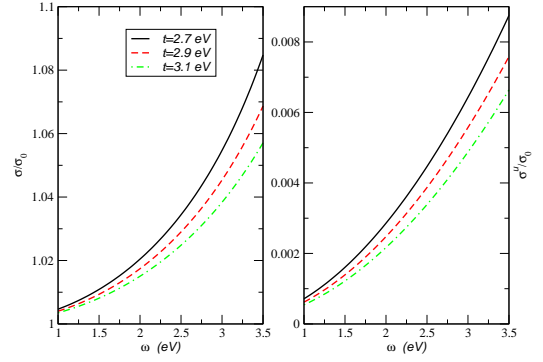}
\caption{(color online) Left: $\sigma(\omega)/\sigma_0$ as a function
of the frequency, including both Eq. (\ref{Eq_s}) and the
correction $\Re\sigma^u_{xx}$, Eq. (\ref{uglySigma}), for several values of $t$.
Right: The correction   $\Re\sigma^u_{xx}$, given by Eq. (\ref{uglySigma}), for several values of $t$.
It is clear that the contribution from this term has barely no effect on the results given by
Eq. (\ref{Eq_s}). The calculations are for zero chemical potential and for room temperature (there is no visible effect on $\sigma(\omega)/\sigma_0$ in the visible range of the spectrum, when compared
to a zero temperature calculation).
\label{ss0}}
\end{center}
\end{figure}

We now want to make quantitative the effect of the term given by
Eq. (\ref{Eq_dir}), which was neglected in Eq. (\ref{Eq_s}). To that end we expand the function $\phi(\bm k)$
up to third order in momentum. The expansion is
\begin{eqnarray}
\phi(\bm k)&\simeq& \frac {3a}2(k_y-ik_x)+
\frac 1 2 \left(\frac {3a}2\right)^2(k^2_x+k^2_y/3+2ik_xk_y)
\nonumber\\
&+&\frac 1 6  \left(\frac {3a}2\right)^3
(ik_x^3-k^3_y/3-3k^2_xk_y+ik^2_yk_x)\;.
\end{eqnarray}
The angular integral in  Eq. (\ref{Eq_dir}) leads to

\begin{equation}
\int^{2\pi}_0d\theta\left(
[\Re\phi(\bm k)]^2-[\Im\phi(\bm k)]^2
\right)=\frac {\pi}{24}\left(\frac {3ak}2\right)^4\,,
\end{equation}  
where we still assume  $\vert\phi(\bm k)\vert=3ak/2$. Within this 
approximation the contribution to the conductivity coming from
Eq. (\ref{Eq_dir}) has the form
\begin{equation}
\label{uglySigma}
\Re\sigma_{xx}^u(\omega)=\sigma_0\frac 1 {4! 2^4}\left(
\frac {\hbar\omega}t
\right)^2\left(
\tanh\frac {\hbar\omega+2\mu}{4k_BT}+
\tanh\frac {\hbar\omega-2\mu}{4k_BT}
\right)\,.
\end{equation}
Due to the prefactor, this contribution has only a small effect and
shows that the current operator basically conserves the circular
symmetry found close to the $K$-points. In Figure \ref{ss0} we present
$\sigma(\omega)/\sigma_0$ as function of the frequency, considering
several values of $t$, in the optical range and also discussing the numerical value of the term given in 
 Eq. (\ref{uglySigma}).

\subsection{The imaginary part of the conductivity} 

Neglecting the term proportional to Eq. (\ref{Neglect}), the imaginary
part of the conductivity is given by
\begin{eqnarray}
\Im\sigma_{xx}(\omega)&=&\frac 1{\hbar\omega}\frac 4 {\pi}\sigma_0
(\mu-\frac 2 9 \mu^3/t^2)
-\frac{\sigma_0}{\pi}
\log\frac {\vert \hbar\omega+2\mu\vert}{\vert\hbar\omega-2\mu \vert}
\nonumber\\
&-&\frac {\sigma_0}{36\pi}\left(\frac {\hbar\omega}t\right)^2
\log\frac {\vert \hbar\omega+2\mu\vert}{\vert\hbar\omega-2\mu \vert}\,,
\end{eqnarray}
where we have included all the terms that diverge at
$\hbar\omega=2\mu$ and the contribution from the cubic term in
frequency in the density of states. The contribution of the last term
of $f[\phi(\bm k)]$ in (\ref{FormFactor}) is given by
\begin{equation}
\Im\sigma_{xx}^u(\omega)=-\frac {\sigma_0}{18\pi}\frac 1 {4! 2^4}\left(
\frac {\hbar\omega}t
\right)^2
\log\frac {\vert \hbar\omega+2\mu\vert}{\vert\hbar\omega-2\mu \vert}\,.
\end{equation}
 
If we neglect the terms in $\mu^3$ and $\omega^2$ we obtain the same
expressions as those derived by Falkovsky and Pershoguba
\cite{Falkovsky08}. We note that these terms are also obtained from the
polarizability in the limit $q\rightarrow0$ since the Fermi velocity
is not $k$-dependent.\cite{Wunsch06}

\subsection{The Drude weight and the Hall coefficient}

The Drude  weight (or  charge stiffness)
defined by Eq. (\ref{DW}) can be computed in 
different limits. In the case $\mu=0$ we are interested in its
temperature dependence. For zero temperature the
exact relation
\begin{equation}
\sum_{\bm k}\vert\phi(\bm k)\vert=\frac 1 8 \sum_{\bm k}
\frac {f[\phi(\bm k)]}{\vert\phi(\bm k)\vert}
\end{equation}
assures that $D=0$ when $\mu=0$.
In general, the Drude weight has the following form:
\begin{eqnarray}
D(T,\mu)&=&t\sigma_0\frac {4\pi^2}{3\sqrt 3}
\frac 1 {N_c}\sum_{\bm k}\left[
\vert\phi(\bm k)\vert - \frac 1 8
\frac {f[\phi(\bm k)]}{\vert\phi(\bm k)\vert}
\right]\nonumber\\
&\times&\left[\tanh\frac{t\vert\phi(\bm k)\vert +\mu}{2k_BT}
+\tanh\frac{t\vert\phi(\bm k)\vert -\mu}{2k_BT}
\right]\,.
\end{eqnarray}
In the case of finite $\mu$, the temperature dependence of $D(T,\mu)$
is negligible. In the Dirac cone approximation we obtain
\begin{equation}
D(0,\mu)=4\pi \sigma_0\mu\left[
1 -\frac 1 9 \left(\frac {\mu}t
\right)^2
\right]\,.
\end{equation}
On the other hand, at zero chemical potential the temperature dependence
of the charge stiffness is given by
\begin{equation}
D(T,0)=8\pi\ln 2\sigma_0 k_BT -4\pi\zeta(3)\sigma_0\frac{(k_BT)^3}{t^2}\,,
\end{equation}
where $\zeta(x)$ is the Riemann zeta function. 

Zotos {\it et al.} have shown a very general relation between the Drude
weight and the Hall coefficient.\cite{Zotos1,Zotos2} This relation is
\begin{equation}
R_H=-\frac {1}{eD}\frac{\partial D}{\partial n}\,.
\label{RH}
\end{equation}  
Equation (\ref{RH}) does not take into account the possibility of valley
degeneracy and therefore it has to be multiplied by two when we apply it
to graphene. In the case of a finite chemical potential we have the
following relations between the Fermi wave vector $k_F$ and the 
chemical potential: $n=k_F^2/\pi$ and $\mu=2tak_F/3$. Applying 
Eq. (\ref{RH}) to graphene we obtain
\begin{equation}
R_H=-\frac 2 e \frac {n^{-1/2}/2-3a^2\sqrt n/8}
{\sqrt n-a^2n^{3/2}/4}\simeq -\frac 1{en}\,.
\end{equation}
\section{Effect of $t'$ on the conductivity of graphene} 
\label{stpprime}

In this section we want to discuss the effect of $t'$ on the conductivity
of graphene. One important question is what the value of $t'$ is in graphene.
Deacon {\it et al.}\cite{Deacon} 
proposed that the dispersion for graphene, obtained from a tight-binding approach with non-orthogonal basis functions, is  of the form
\begin{equation}
E=\pm \frac{t\vert\phi(\bm k)\vert}{1\mp s_0\vert\phi(\bm k)\vert}\, 
\label{Eq_decon}
\end{equation} 
with $\vert\phi(\bm k)\vert\simeq \frac 3 2 ka$, with $a$ the carbon-carbon
distance. On the other hand the dispersion of graphene including
$t'$ has the form

\begin{equation}
E= \pm t \frac 3 2 ka - t'\left[ \frac 9 4(ka)^2-3\right]\,.
\end{equation}
To relate $t'$ and $s_0$ we expand Eq. (\ref{Eq_decon}) as
\begin{equation}
E\simeq \pm t \vert\phi(\bm k)\vert (1\pm s_0\vert\phi(\bm k)\vert)
=\pm t\frac 3 2 ka+s_0t\frac 9 4(ka)^2 \,,
\end{equation} 
which leads $t'/t=-s_0$ with $s_0=0.13$. 

For computing the conductivity of graphene we need to know the Green's
function with $t'$. These can be written in matrix form as

\begin{eqnarray}  
\label{eq:funcao2_green_rede}
  \trG^0(\bm k,i\omega_n) &=&  \sum_{\alpha = +,-} 
  \frac{1/2}
  { i\omega_n - \alpha t\vert\phi(\bm k)\vert/\hbar +
  2t'[\vert\phi(\bm k)\vert^2-3]/\hbar }\nonumber\\
&\times&\left(  
\begin{array}{cc}
   1 & -\alpha \phi(\bm k)/\vert\phi(\bm k)\vert  \\
   -\alpha \phi(\bm k)^\ast/\vert\phi(\bm k)\vert & 1
  \end{array}
\right),
\label{green}
\end{eqnarray}
where $\trG^0(\bm k,i\omega_n)$ stands for 
\begin{equation}
\trG^0(\bm k,i\omega_n)=
\left(
\begin{array}{cc}
G_{AA}(\bm k,i\omega_n) & G_{AB}(\bm k,i\omega_n)\\
G_{BA}(\bm k,i\omega_n) & G_{BB}(\bm k,i\omega_n)
\end{array}
\right)\,.
\end{equation}
From Eq. (\ref{green}) we see that only the poles are modified, with the
coherence factors having the same form as in the case with $t'=0$.
The current operator $j_x^P=j_{x,t}^P+j_{x,t'}^P$, as derived from the tight-binding Hamiltonian
is written in momentum space as

\begin{eqnarray}
j_{x,t}^P&=&\frac {tiea}{2\hbar}\sum_\sigma\sum_{\bm k}
[(\phi(\bm k)-3)a^\dag_{\sigma,\bm k}b_{\sigma,\bm k}-\nonumber\\
&-&(\phi^\ast(\bm k)-3)b^\dag_{\sigma,\bm k}
a_{\sigma,\bm k}]\,,
\end{eqnarray}
and
\begin{eqnarray}
j_{x,t'}^P&=&\
\frac {3t'iea}{2\hbar}\sum_\sigma\sum_{\bm k}[\phi(\bm k)-\phi^\ast(\bm k)]
\times\nonumber\\
&&(a^\dag_{\sigma,\bm k}a_{\sigma,\bm k}+
b^\dag_{\sigma,\bm k}b_{\sigma,\bm k})\,.
\end{eqnarray}
The operators $j_{x,t}^P$ and $j_{x,t'}^P$ are the current operators
associated with the hopping amplitudes $t$ and $t'$, respectively. The
current-current correlation function is now a sum of three different
terms: one where we have two $j_{x,t}^P$ operators, another one where
we have a $j_{x,t}^P$ and a $j_{x,t'}^P$, and a third one with two
$j_{x,t'}^P$. This last term vanishes exactly, since it would
correspond to the current-current correlation function of a triangular
lattice. Also the crossed term vanishes exactly which can be
understood by performing a local gauge transformation to the fermionic
operators of one sub-lattice, only. The first term leads to a
contribution of the same form as in Eq. (\ref{Eq_s}) but with the
numerators of the two $\tanh$ replaced by
$E_+=\hbar\omega+2t'[(\hbar\omega)^2/(4t^2)-3]+2\mu$ and
$E_-=\hbar\omega-2t'[(\hbar\omega)^2/(4t^2)-3]-2\mu$, respectively.

\begin{figure}[t]
\begin{center}
\includegraphics*[angle=0,width=0.8\linewidth]{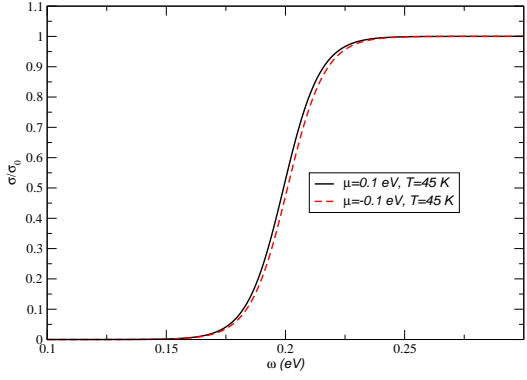}
\caption{(color online) Real part of the conductivity for two values of the chemical potential
at the temperature of 45 K. The parameters used are
$t=3.1$ eV and $t'=-0.13t$. Only the energy range of $\omega\in[0.1,0.3]$ eV is shown because
only here has the chemical potential difference any noticeable effect.
\label{condtprime}}
\end{center}
\end{figure}

As a consequence the effect of
$t'$ in the conductivity a graphene only enters in the band structure
$E_{\pm}$ in the Fermi functions. In Figure \ref{condtprime} we plot
the real part of the optical conductivity for two different values of
$\mu$, one with the Fermi energy in the conduction band and the other with
the Fermi energy in the valence band. There is a small effect near twice
the absolute value of the chemical potential, due to the breaking of
particle-hole symmetry introduced by $t'$. For optical frequencies, the effect of $t'$ is negligible.

\section{The electromagnetic scattering problem}
\label{scattering}
\begin{figure}[t]
\begin{center}
\includegraphics*[angle=0,width=0.8\linewidth]{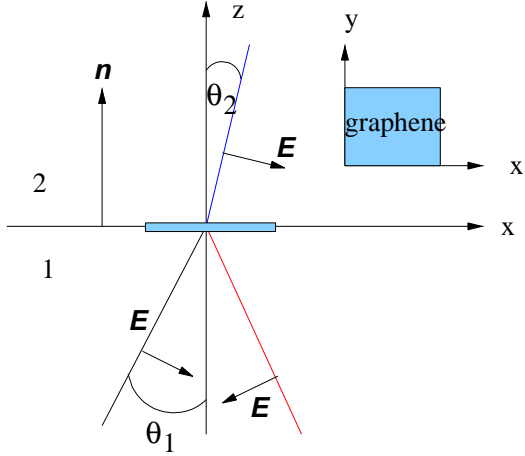}
\caption{(color online) Geometry of $p$ polarized light scattering between two media with graphene
separating them. The electrical permittivities of the two media
are $\epsilon_i\epsilon_0$, with $i=1,2$
\label{fig2}}
\end{center}
\end{figure}

Here we derive the reflectivity and the transmissivity of light
between two media, characterized by
electrical permittivities $\epsilon_i\epsilon_0$, with $i=1,2$,
 separated by a graphene flake. The scattering
geometry is represented in Fig. \ref{fig2}, i.e., we assume the field 
to propagate in the direction $\bm k=(k_x,0,k_z)$. 

In the following, we assume the field to be given by $\bm
E=(E_x,0,E_z)$ ($p$ polarization). The case of $s$
polarization is addressed in Appendix \ref{app:transmissivity}.

The electromagnetic boundary conditions then are\cite{Jackson}
\begin{eqnarray}
(\bm D_2-\bm D_1)\cdot\bm n=\rho\,,\\
\bm n\times(\bm E_2-\bm E_1)=0\,,
\end{eqnarray}
where $\rho$ is the surface charge density, in our case the
graphene charge density. If we represent the intensity
of the incident, reflected, and transmitted electric field as
$E_i$, $E_r$, and $E_t$, respectively, the boundary conditions
can be written as
\begin{eqnarray}
\label{BCone}
(E_i-E_r)\cos\theta_1=E_t\cos\theta_2\,,\\
-\epsilon_2\epsilon_0E_t\sin\theta_2+
\epsilon_1\epsilon_0(E_i+E_r)\sin\theta_1=\rho\,,
\label{BCtwo}
\end{eqnarray}
where $\epsilon_0$ is the vacuum permittivity, $\epsilon_1$ and
$\epsilon_2$ are the relative permittivity of the two media and
$\theta_1$ and $\theta_2$ are the incident and refracted angle,
respectively.  Now the continuity equation in momentum space reads
\begin{equation}
\rho(\omega)=j_x(\omega)k_x/\omega\,,
\label{continuity}
\end{equation}
and Ohm's law is
written as
\begin{equation}
j_x(\omega)=\sigma(\omega)E_x=\sigma(\omega)E_t\cos\theta_2\,.
\label{ohm}
\end{equation}
Combining Eqs. (\ref{BCone}) - (\ref{ohm}),
 we arrive at the following result, valid for normal
incidence, for the  transmissivity $T$
\begin{equation}
T=\sqrt{\frac{\epsilon_2}{\epsilon_1}}
\frac {4(\epsilon_1\epsilon_0)^2}{|(\sqrt{\epsilon_1\epsilon_2}
+\epsilon_1)\epsilon_0+
\sqrt{\epsilon_1}\sigma(\omega)/c|^2}\,.
\end{equation}
If we now consider both media to be vacuum and that the graphene
is at half filling ($\sigma(\omega)\simeq\sigma_0$) we obtain
\begin{equation}
T=\frac 1 {(1+\pi\alpha/2)^2}\simeq 1-\pi\alpha\,,
\end{equation}
where $\alpha=e^2/(4\pi\epsilon_0c\hbar)$, is the fine structure constant. The reflectivity is also
controlled by the fine structure constant $\alpha$. For normal incidence
it reads
\begin{equation}
R=\frac {|\sqrt{\epsilon_1\epsilon_2}\epsilon_0+\sqrt{\epsilon_1}
\sigma(\omega)/c-\epsilon_1\epsilon_0|^2}{|\sqrt{\epsilon_1\epsilon_2}\epsilon_0+\sqrt{\epsilon_1}
\sigma(\omega)/c+\epsilon_1\epsilon_0|^2}\,,
\end{equation}
and if both media are the vacuum we obtain
\begin{equation}
R=\frac {\pi^2\alpha^2}4T\,.
\end{equation}

\begin{figure}[t]
\begin{center}
  \includegraphics*[angle=0,width=0.8\linewidth]{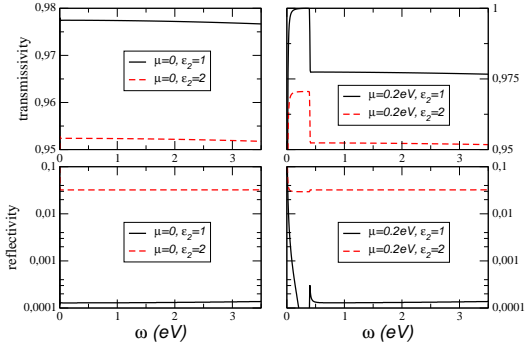}
\caption{(color online) The transmissivity and reflectivity for normal incident as function
of the frequency for $T=10K$ where the first medium is vacuum ($\epsilon_1=1$) and the second medium is either vacuum ($\epsilon_2=1$) or a SiO${}_2$-substrate ($\epsilon_2=\epsilon_\infty=2$). Left: At zero chemical potential. Right: At finite chemical potential $\mu=0.2$eV.
\label{transmissivity}}
\end{center}
\end{figure}

In Fig. \ref{transmissivity}, the transmission and reflection
coefficients for normal incident as function of the frequency for
temperature $T=10K$ are shown where the first medium is vacuum
($\epsilon_1=1$) and the second medium is either vacuum
($\epsilon_2=1$) or a SiO${}_2$-substrate
($\epsilon_2=\epsilon_\infty=2$, $\epsilon_\infty$ being the
high-frequency dielectric constant of SiO${}_2$). The left hand side
shows the data for zero doping and the right hand side for finite
doping $\mu=0.2$eV. In Appendix \ref{app:transmissivity}, we present
the formulas for arbitrary angle of incidence.

It is interesting to compare the result for graphene with that for
bilayer graphene.  For the bilayer, the transmissivity is given by
\cite{Falko07}

$$
T = 1-2\pi\alpha f_2(\omega)
$$

with $f_2(\omega)$ given by

\begin{eqnarray}
f_2(\omega)&=&\frac {\hbar\omega+2t_\perp}{2(\hbar\omega+t_\perp)}
+\frac{\theta(\hbar\omega-t_\perp)}{(\hbar\omega/t_\perp)^2}
\nonumber\\
&+&
\frac{(\hbar\omega-2t_\perp)\theta(\hbar\omega-2t_\perp)}{2(\hbar\omega-t_\perp)}\,,
\end{eqnarray}
and $t_\perp$ the hopping amplitude between the graphene planes.  For
frequencies much larger than $t_\perp$, which is the case in an
experiment done in the visible region of the spectrum, one obtains

\begin{equation}
f_2(\omega)\simeq 1 - \frac {t_\perp^2}{(\hbar\omega)^2}\simeq 1\,,
\end{equation}
which leads to $T \simeq 1-2\pi\alpha$. Again, as in graphene, the transmissivity is controlled
by the fine structure constant. It is interesting to note that for $\hbar\omega\ll t_\perp$
we also obtain the same result for $T$. 

The appearance of the fine structure constant $\alpha$ in the two cases is connected to the spinorial structure of the electronic wave function. In other words, the reduction of the transmissivity through a clean system is caused by a universal current induced by interband transitions.

\section{Conclusions}
\label{Concl}
We have presented a detailed study of the optical properties of graphene based on the general, non-interacting tight-binding model. Special emphasis was placed on going beyond the usual Dirac-cone approximation, i.e., we included the cubic term in the density-of-states. The conductivity was thus consistently calculated to order $(\hbar\omega/t)^2$ for arbitrary chemical potential and temperature.  

We also assessed the effect of the next nearest neighbor coupling $t'$ on the optical properties. We find that the additional terms to the current operator do not contribute to the conductivity and that modifications only enter through the modified energy dispersion.

Using the full conductivity of clean graphene, we determine the transmissivity and reflectivity of light that is scattered from two media with different permittivity and graphene at the interface. Our results are important for optical experiments in the visible frequency range.\cite{nair} For example, the apparent disagreement between the presented theory for graphene and experiments by Dawlaty {\it et al.}\cite{George} at visible frequencies indicates that the interlayer interaction in epitaxial-SiC graphene is significant and cannot be neglected.

\section*{Acknowledgements}
This work was supported by the ESF Science Programme INSTANS
 2005-2010, and by FCT under the grant PTDC/FIS/64404/2006.
\appendix
\section{Eq. (\ref{Eq_dir}) up to first order in momentum}
\label{app:Neglect}
The function $\phi(\bm k)$ is given close to the Dirac point
by
\begin{eqnarray}
\phi(\bm k)\simeq \frac {3a} 2 (k_y-ik_x)\,.
\end{eqnarray}
This leads to the following result
\begin{eqnarray}
T(\theta)=
\frac {[\Re \phi(\bm k)]^2-[\Im \phi(\bm k)]^2 }{\vert \phi(\bm k)\vert^2}
=-\cos(2\theta)
\end{eqnarray}

It is now easy to see that
\begin{equation}
\int_0^{2\pi}d\theta T(\theta)g(\vert\phi(\bm k)\vert)=0,
\end{equation}
where we have used the result $\vert\phi(\bm k)\vert=3ak/2$,
valid near the Dirac point.

\section{Transmissivity and reflectivity for arbitrary incidence}
\label{app:transmissivity}
Here we present the general formula for the transmissivity and
reflectivity of light being scattered at a plane surface between two
media of different dielectric properties and a graphene sheet at the
interface.

For $p$ polarization, the reflection and transmission amplitude are obtained from the boundary conditions of Eqs. (\ref{BCone}), (\ref{BCtwo}) and read
\begin{align}
r&=\frac{M-1}{M+1}\;,\;t=\sqrt{\frac{\epsilon_1}{\epsilon_2}}\frac{2K}{M+1}
\end{align}
with $M=K+\Sigma\cos\theta_1$, where $\theta_1$ denotes the incident angle and
\begin{align}
K&=\frac{\epsilon_2}{\epsilon_1}\frac{k_z^i}{k_z^t}\;,\;\Sigma=\frac{\sigma(\omega)}{\sqrt{\epsilon_1}\epsilon_0c}\;.
\end{align} 
Above, $k_z^i=\sqrt{\epsilon_1(\omega/c)^2-k_x^2}$ ($k_z^t=\sqrt{\epsilon_2(\omega/c)^2-k_x^2}$) denotes the perpendicular component of the incident (transmitted) wave vector relative to the interface, $k_x$ the parallel (conserved) component and $\epsilon_1$ ($\epsilon_2$) is the dielectric constant of the first (second) medium, see Fig. \ref{fig2}. For $s$ polarization, $r$ and $t$ are independent of the angle of incident and in the Dirac cone approximation yield the same result as for $p$ polarization in the case of normal incident ($\theta_1=0$).

Generally, the reflection and the transmission coefficient are given by $R=|r|^2$ and $T=|t|^2k_z^t/k_z^i$, respectively. For a simple (non-conducting) interface, this leads to the conservation law $T+R=1$. Notice that there is no such conservation in the present case due to absorption within the graphene sheet.

For a suspended graphene sheet with $\epsilon_1=\epsilon_2=1$ at the Dirac point ($\sigma(\omega)\simeq\sigma_0$), the reflection and transmission coefficient for $p$ polarization read
\begin{align}
R&=\frac{(\tilde\alpha\cos\theta_1)^2}{(1+\tilde\alpha\cos\theta_1)^2}\;,\;T=\frac{1}{(1+\tilde\alpha\cos\theta_1)^2}\;,
\end{align}
with $\tilde\alpha=\pi\alpha/2$ and $\alpha=e^2/(4\pi\epsilon_0 c\hbar)$ the fine structure constant.


\end{document}